\documentclass{aa}

\usepackage{psfig}
\usepackage{graphicx}
\usepackage{natbib}
\usepackage[english]{babel}  
\usepackage{times}

\def\gsim{\;\lower4pt\hbox{${\buildrel\displaystyle >\over\sim}$}\,}

\begin{document}

\title{Generation of radiative knots in a randomly pulsed protostellar
jet}

\subtitle{I. Dynamics and energetics}

\author{R. Bonito\inst{1, 2} \and S. Orlando\inst{2} \and G. Peres\inst{1, 2}
\and J. Eisl\"offel\inst{3} \and M. Miceli\inst{1, 2} \and F. Favata\inst{4}}

\offprints{R. Bonito\\ \email{sbonito@astropa.unipa.it}}

\institute{Dip. Scienze Fisiche ed Astronomiche, Sez. Astronomia,
Universit\`a di Palermo, P.zza del Parlamento 1, 90134
Palermo, Italy
\and 
INAF -- Osservatorio Astronomico di Palermo, P.zza del Parlamento 1,
90134 Palermo, Italy 
\and
Th\"uringer Landessternwarte, Sternwarte 5, D-07778 Tautenburg, Germany
\and
European Space Agency
Community Coordination and Planning Office
8-10 rue Mario Nikis
F-75738 Paris cedex 15
France
} 

\date{Received, accepted}

\authorrunning{}
\titlerunning{}

\abstract
{Herbig-Haro objects are characterized by a complex knotty morphology
detected mainly along the axis of protostellar jets in a wide range of bands:
from radio to IR to optical bands, with X-rays knots also detected in
the last few years. Evidence of interactions between knots formed
in different epochs have been found, suggesting that jets may result
from the ejection of plasma blobs from the stellar source.}
{We aim at investigating the physical mechanism leading to the
irregular knotty structure observed in protostellar jets in different
wavelength bands and the complex interactions occurring among blobs of
plasma ejected from the stellar source.}
{We perform 2D axisymmetric hydrodynamic numerical simulations
of a randomly ejected pulsed jet. The jet consists of a train of
blobs which ram with supersonic speed into the ambient medium. The initial
random velocity of each blob follows an exponential distribution. We explore
the ejection rate parameter in order to derive constraints on
the physical properties of protostellar jets by comparison of
model results with observations. Our model takes into account
the effects of radiative losses and thermal conduction.}
{We find that the mutual interactions of blobs ejected at different
epochs and with different speed lead to a variety of plasma components
not described by current models of jets. The main features characterizing
the random pulsed jet scenario are: single high speed knots, showing
a measurable proper motion in nice agreement with optical and X-rays
observations; irregular chains of knots aligned along the jet axis and
possibly interacting with each other; reverse shocks interacting with
outgoing knots; oblique shock patterns produced by
the reflection of shocks at the cocoon surrounding the jet. All these
structures concur to determine the morphology of the jet in different
wavelength bands. We also find that the thermal conduction plays
a crucial role in damping out hydrodynamic instabilities that would
develop within the cocoon and that contribute to the jet breaking.}
{}


\maketitle

\section{Introduction}
\label{Introduction}

Herbig-Haro (HH) objects characterize the emission regions near protostars
where the accretion process is still ongoing, often allowing us to
identify the presence of a young stellar object (YSO) not observable
otherwise. The HH objects are observed in all wavelength bands (from
radio and IR to optical and UV, and also in the X-rays; see the review of
\citealt{rb01}) as a chain of knots within the supersonic protostellar
jets. The knotty structure has been frequently observed in the
optical band in many HH objects and has been discovered for the first time
also in X-rays in HH 154 (\citealt{fbm06}, \citealt{bff08}), one among the
nearest and most luminous X-ray emitting protostellar jets. The HH 154 jet
is one of the best studied, being observed through multi-wavelength and
multi-epoch observations with several instruments, among them the Hubble
Space Telescope (HST; \citealt{fl98}; \citealt{bff08}), Chandra/ACIS-I
(\citealt{bfr03}; \citealt{fbm06}), and XMM-Newton (\citealt{ffm02}).

The origin of chains of optical knots commonly observed in jets have
been traditionally interpreted through models of jet ejected with a
sinusoidal variable velocity (e.g. \citealt{rdk07}). In fact, in these
models, the internal shocks resulting from plasma flowing with different
velocity lead to a nodular structure, constituted by a regular chain of
knots with individual proper motions. An alternative scenario
has been proposed by \citet{yfc09} who model, through 3D simulations, the appearing of clumps
within the jet beam (with random locations, sizes, and speeds), and
explore the possibility of intrinsic density heterogeneity of the jet.
Original models of jets accounting for stochastic jets and random
time variability have been previously discussed by \citet{rob86} (in
the context of extragalactic jets) and \citet{rag92} (in the limit of
large distances from the stellar source). Also \citet{vjo02} suggested
a disk/jet launching model with a combination of a continuous and a random
component of the velocity, thus justifying the random speed models.

Recently, multi-wavelength observations of HH 154 collected in 2005 with
both Chandra and HST showed evidence of mutual interactions between
bright knots, formed probably at different epochs and with different
speeds (\citealt{bff08}). These features seem to be common also in other
objects (\citealt{gef09}) and are not predicted in models of jets with
sinusoidal velocity. Moreover as discussed by \citet{yfc09},
the observed knots could be interpreted erroneously as the product of
a sinusoidal variable velocity jet.

The aim of this work is to investigate, through detailed numerical
modeling, the origin and dynamics of the irregular knotty structure
observed in several bands (and, in particular, in X-rays) in
protostellar jets and the possible interactions between knots. Moreover,
we aim at deriving general model predictions to be compared with
observations of X-ray emitting protostellar jets (DG Tau, HH 2, etc.) to
understand the physical mechanisms leading to X-ray emission in
different objects.

In our previous studies, we have investigated the origin of X-ray
emission from protostellar jets, through a model of a continuously ejected
supersonic jet (\citealt{bop04}, \citealt{bop07}). Such a model predicts
the propagation of a compact X-ray source through the circumstellar
medium with spectral behavior, X-ray luminosity, and proper motion in
good agreement with the observations (\citealt{ffm02}). However, the
model does not reproduce the complex knotty morphology and variability
of the X-ray source detected in HH 154 (\citealt{fbm06}).

Here we suggest an improvement to our previous model by considering the
scenario of a pulsed jet interacting with an initially homogeneous ambient
medium. The model presented here describes a jet constituted by a train
of plasma blobs randomly ejected by the stellar source. The initial
velocity of our model follows an exponential distribution and this
choice corresponds to a distribution of many low-speed blobs and few blobs
with high initial ejection velocity. The blobs are expected to interact
with each other leading to the formation of knots with individual proper
motions, possibly observable in various bands.  In the following, we will
refer to blobs to consider the plasma ejected from the stellar source
and to knots to consider the observable structures within the jet.

We plan to present our results in two papers. In the present paper
we describe the dynamics and energetic of the pulsed jet/ambient
interaction, exploring the influence of the ejection rate of plasma
blobs. In the second paper (Bonito et al. 2009, in preparation) we will focus
on the variability and morphology of X-ray emitting features predicted
by the model as they would be observed with Chandra/ACIS-I and with
XMM/Newton.

The paper is organized as follows: in Sect. \ref{The model} we describe the
numerical model of the pulsed jet, its initial setup, and the exploration
of the parameter space; in Sect. \ref{Results} we analyze the hydrodynamic
evolution of the model, focusing on the internal structures occurring
within the jet, and the physical effects due to the thermal conduction
process; in Sect. \ref{Discussion} we discuss the results and their
physical interpretation, and  in Sect. \ref{Conclusions} we draw our
conclusions.

\section{The model}
\label{The model}

The model describes a protostellar jet ramming with a supersonic
variable ejection speed into an initially homogeneous ambient medium.
The fluid is assumed to be fully ionized\footnote{See \citealt{bop07}.} with a ratio of specific heats
$\gamma = 5/3$. The model takes into account the radiative cooling
from optically thin plasma, and the thermal conduction (including the
effects of heat flux saturation). The propagation of the jet through the
surrounding medium is modeled by solving numerically the time-dependent
hydrodynamic equations:

\begin{equation}
\frac{\partial \rho}{\partial t} + \nabla \cdot \rho \mbox{\bf v} = 0
\label{eq:massa}
\end{equation}

\begin{equation}
\frac{\partial \rho \mbox{\bf v}}{\partial t} +\nabla \cdot \rho
\mbox{\bf vv} + \nabla p = 0
\label{eq:momento}
\end{equation}

\begin{equation}
\frac{\partial \rho E}{\partial t} +\nabla\cdot (\rho E+p)\mbox{\bf v}
= - \nabla\cdot q - n_e n_H P(T)
\label{eq:en+r+c}
\end{equation}

\noindent
with

\[
p = (\gamma-1)\rho\epsilon~,~~~~~~~~~~~~~
E = \epsilon +\frac{1}{2} |\mbox{\bf v}|^2 ~,
\]

\noindent
where $p$ is the pressure, $E$ the total gas specific energy
(internal energy, $\epsilon$, and kinetic energy) respectively, $\rho$
is the mass density, $t$ the time, $\bf v$ the plasma velocity, $q$
the heat flux, $n_e$ and $n_H$ are the electron and hydrogen density,
respectively, $P(T)$ is the radiative losses function per unit emission
measure\footnote{$P(T)$ is described by a functional form that takes
into account free-free, bound-free, bound-bound and 2 photons emission
(see \citealt{rs77}; \citealt{mgv85}; \citealt{km00}).}, $T$ the plasma
temperature.

Following \citet{db93}, we use an interpolation expression for the thermal
conductive flux which allows for a smooth transition between the Spitzer
(\citealt{spi62}) and saturated (\citealt{cm77}) conduction regimes

\begin{equation}
q = \left(\frac{1}{q_{\rm spi}}+\frac{1}{q_{\rm sat}}\right)^{-1} ~,
\label{eq:flusso} 
\end{equation}

\noindent
where

\begin{equation}
q_{\rm spi} = - \kappa (T) \nabla T ~,
\label{eq:flussospi}
\end{equation}

\noindent
where $\kappa (T) = 9.2\times10^{-7} T^{5/2}$ erg s$^{-1}$ K$^{-1}$
cm$^{-1}$ is the thermal conductivity, and

\begin{equation}
q_{\rm sat} = - sign(\nabla T) 5 \phi \rho c_{\rm s}^{3} ~,
\label{eq:flussosat}
\end{equation}

\noindent
where $\phi\sim0.3$ (\citealt{1984ApJ...277..605G};
\citealt{1989ApJ...336..979B}, and references therein) and $c_{\rm s}$
is the isothermal sound speed.

The hydrodynamic equations are solved using the FLASH code
(\citealt{for00}) with customized numerical modules for the optically
thin radiative losses and the thermal conduction (see \citealt{opr05},
for the details of the implementation of the thermal conduction). The
FLASH code uses a directionally split Piecewise Parabolic Method (PPM)
solver to handle compressible flows with shocks (\citealt{cw84}), the
PARAMESH library to handle adaptive mesh refinement (\citealt{mom00}), and
the Message-Passing Interface library to achieve parallelization.

\subsection{Numerical setup}
\label{Numerical setup}

We solve the hydrodynamic equations in 2D, using cylindrical
coordinates in the plane ($r, z$), and assuming axisymmetry with the
jet axis coincident with the axis of symmetry ($z$). The computational
grid size is (2000 AU $\times$ 6000 AU). The maximum spatial resolution
achieved by our simulations is $\approx 8$ AU, as determined with the
algorithm of the PARAMESH library for $4$ refinement levels, which
corresponds to covering the initial jet radius with $4$ points at the
maximum resolution \footnote{The maximum resolution corresponds to $\approx 8$
times the resolution of Chandra/ACIS-I observations and to about twice the spatial resolution of HST, 0.4''$\approx
60$ AU at 150 pc and 0.1''$\approx 15$ AU at 150 pc, i.e. at the distance of the nearest star forming region in Taurus.}.

\begin{figure}[!t]
\centerline{\psfig{figure=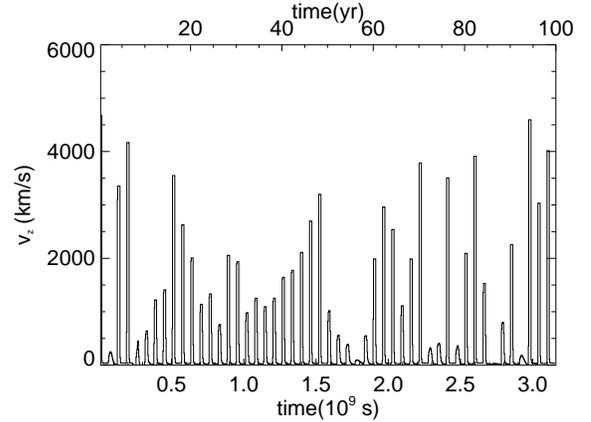,width=8cm}}
\caption{Initial jet velocity of each blob as a function of time for
the $\Delta t$ = 2 yr case (run LJ2).}
\label{vely}
\end{figure}

We follow the evolution of the jet for a time interval ranging
between 100 and 400 yrs, depending on the set of model parameters.
The pulsed jet consists of a train of blobs, each lasting for $0.5$ yr, with an ejection rate
corresponding to a time interval between two consecutive blobs $\Delta t$.
Each blob is ejected with a random velocity directed along the
jet axis following an exponential distribution

\begin{equation}
v(r=0,t) = v_{j MAX} (b-log(rand(t)-a)) ~,
\label{eq:v-rand}
\end{equation}

\noindent
where

\begin{equation}
b = log\frac{1}{1-e^{-1}} ~,
\label{eq:b}
\end{equation}

\begin{equation}
a = \frac{e^{-1}}{e^{-1}-1} ~,
\label{eq:a}
\end{equation}

\noindent
and $rand$ is a number ranging between $0$ and $1$ obtained using a
random number generator. The maximum velocity of the blobs is $v_{j MAX}
= 4680$ km/sec and corresponds to the highly supersonic initial Mach
number $M = 1000$. Fig. \ref{vely} shows the initial velocity values of
each blob ejected with an ejection rate corresponding to $\Delta t$ =
2 yr. We generated the random seed once for all the runs so the values
follow always the same sequence, only the rate is different.

The initial random velocity of each blob is along the $z$ axis, coincident
with the jet axis, and its radial profile is

\begin{equation}
v(r,t) = \frac{v(r=0,t)}{\nu \cosh(r/r_j)^{w}-(\nu-1)} ~,
\end{equation}

\noindent
where $v(r=0,t)$ is the on-axis velocity, $\nu$ is the ambient-to-jet density
ratio, $r_{\rm j}$ is the initial jet radius and $w = 4$ is the steepness
parameter for the shear layer which allows a smooth transition of the
kinetic energy at the interface between the jet and the ambient medium.
This layer avoids the growth of random perturbation at the jet border
(\citealt{bmf94}). The density profile as a function of the radial
direction is

\begin{equation}
\rho(r) = \rho_j\left(\nu-\frac{\nu-1}{\cosh(r/r_j)^{w}}\right) ~,
\end{equation}

\noindent
where $\rho_{\rm j}$ is the jet density (see also the velocity and
density profiles as a function of the radial distance from the axis in
\citealt{bop07}, Fig. 1).

Reflection boundary conditions are imposed along the jet axis (consistent
with the adopted symmetry), inflow boundary conditions at the base for
$r < r_{j}$, where $r$ is the radial distance in cylindrical coordinates
and $r_{j}$ is the jet radius, and outflow boundary conditions elsewhere.

\subsection{Parameter space}
\label{Parameters}

The model parameters are derived from the optical and
X-ray data analysis of protostellar jets (e.g. \citealt{pbn06};
\citealt{bff08}; \citealt{fbm06}) and from the results of previous models
(\citealt{bop04,bop07}). Tab. \ref{tab_mod} summarizes the physical
parameters characterizing the simulations.

\begin{table}[!t]
\caption{Summary of the initial physical parameters characterizing the
simulations: $\Delta t$ is the time between two consecutive blobs, $\nu$ is
the ambient-to-jet density contrast, $M$ is the initial jet Mach number,
$v_{\rm j}$ is the velocity of each ejected blob (here we indicate the
range of values randomly generated in our model), $T_{\rm j}$ is the
initial jet temperature, $n_{\rm a}$ is the ambient density. In all the
models we assume: $r_{\rm j} = 30$ AU as initial jet radius, $n_{\rm
j} = 500$ cm$^{-3}$ as initial jet density, and $T_{\rm a} = 10^{3}$
K as initial ambient temperature.}
\label{tab_mod}
\begin{tabular}{lccccccc}
\hline
\hline
Model & $\Delta t$ & $\nu$  & $M$    & $v_{\rm j}$   & $T_{\rm j}$ & $n_{\rm a}$  \\
      & yr         &        &        & [km s$^{-1}$] & [$10^4$ K]  & [cm$^{-3}$]  \\
\hline
LJ0.5   & $0.5$    & $10$   & $1000$ & $10 - 4700$   & $10^{4}$    & $5000$	  \\
LJ2     & $2$      & $10$   & $1000$ & $10 - 4700$   & $10^{4}$    & $5000$	  \\
LJ8     & $8$      & $10$   & $1000$ & $10 - 4700$   & $10^{4}$    & $5000$	  \\	  
LJ2RD   & $2$      & $10$   & $1000$ & $10 - 4700$   & $10^{4}$    & $5000$	  \\
HJ2     & $2$      & $0.1$  & $300$  & $10 - 1400$   & $10^{2}$    & $50$	  \\
\hline
\end{tabular}
\end{table}

The exploration of the parameter space mainly focuses on the
ejection rate of the blobs composing the jet. In all our simulations,
we impose the initial radius and density of the blobs, namely $r_{j}
\approx 30$ AU and $n_{j} = 500$ cm$^{-3}$, respectively. For the sake
of completeness, we explore both the initial light jet scenario (a jet
initially less dense than the unperturbed ambient medium), and the initial
heavy jet scenario (a jet initially denser than the ambient medium);
we consider, therefore, initial ambient-to-jet density ratio $\nu =
n_{a}/n_{j} = 0.1, 10$ in the heavy and light jet cases, respectively. The
initial blob temperature $T_{j}$ ranges between $10^{2}$ and $10^{4}$ K, while the ambient temperature, $T_{a}$, is derived assuming initial pressure equilibrium between the jet and the unperturbed ambient.

The initial jet Mach number values are $M = v_{\rm j}/c_{\rm a}
= 300$ in the heavy jet case and $1000$ in the light jet case, and
corresponds to a maximum velocity of the first blob of about 5000 km/s.
Note that we are interested on the evolution of the pulsed jet after
a transient phase, i.e. after the first ejected blob of plasma has
perturbed the ambient (which is no more homogeneous). We require,
therefore, a high speed of the first blob in order to perturb quickly the
computational domain (6000 AU). The speed of each ejected blob ranges
between about 10 km/s and 5000 km/s during the evolution of the pulsed
jet. It is worth noting that the apparently high values of the maximum
ejection velocity cannot be compared directly to the observable speeds
of the knots within the protostellar jets. In fact we have chosen the
initial speeds which lead, once the interaction between the ejected plasma
and the ambient medium has taken place, to the observed knots speed of
a few hundred km/s (\citealt{fl98}; \citealt{bff08}).

Since the optical knots in HH jets, in general, show time-scale
variability of a few years (see morphological evolution of the HH 154 jet
in about 2 yr discussed in \citealt{bff08}), with new knots appearing at
the base of the jet in about 2 yr (e.g. HH 1 and HH 34, see the movies
discussed in \citealt{har03}), we have chosen an initial ejection rate
parameter corresponding to a time interval $\Delta t$ = 2 yr between two
consecutive blobs. Then we explore this parameter taking into account
a 1/4 value ($\Delta t$ = 0.5 yr) and a $\times 4$ value ($\Delta t$ =
8 yr) in order to constrain the ejection rate from the comparison between
our model predictions and the observations.

\begin{figure}[!t]
\centerline{\psfig{figure=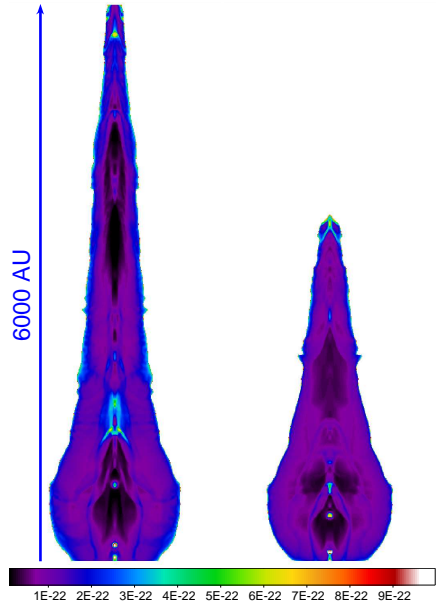,height=8cm}}
\caption{2D spatial distributions of the density of the jet/ambient
interaction in runs LJ2 (left semi-panel) and LJ8 (right semi-panel). The
head of the jet leaves the domain after about 90 yr in LJ2 whereas
in LJ8 it does not reach the end of the domain (6000 AU) in 100 yr.}
\label{dens-2-8}
\end{figure}
 
\section{Results}
\label{Results}

\subsection{Hydrodynamic evolution}
\label{Hydrodynamic evolution}

\begin{figure}[t!]
\centerline{\hbox{
\psfig{figure=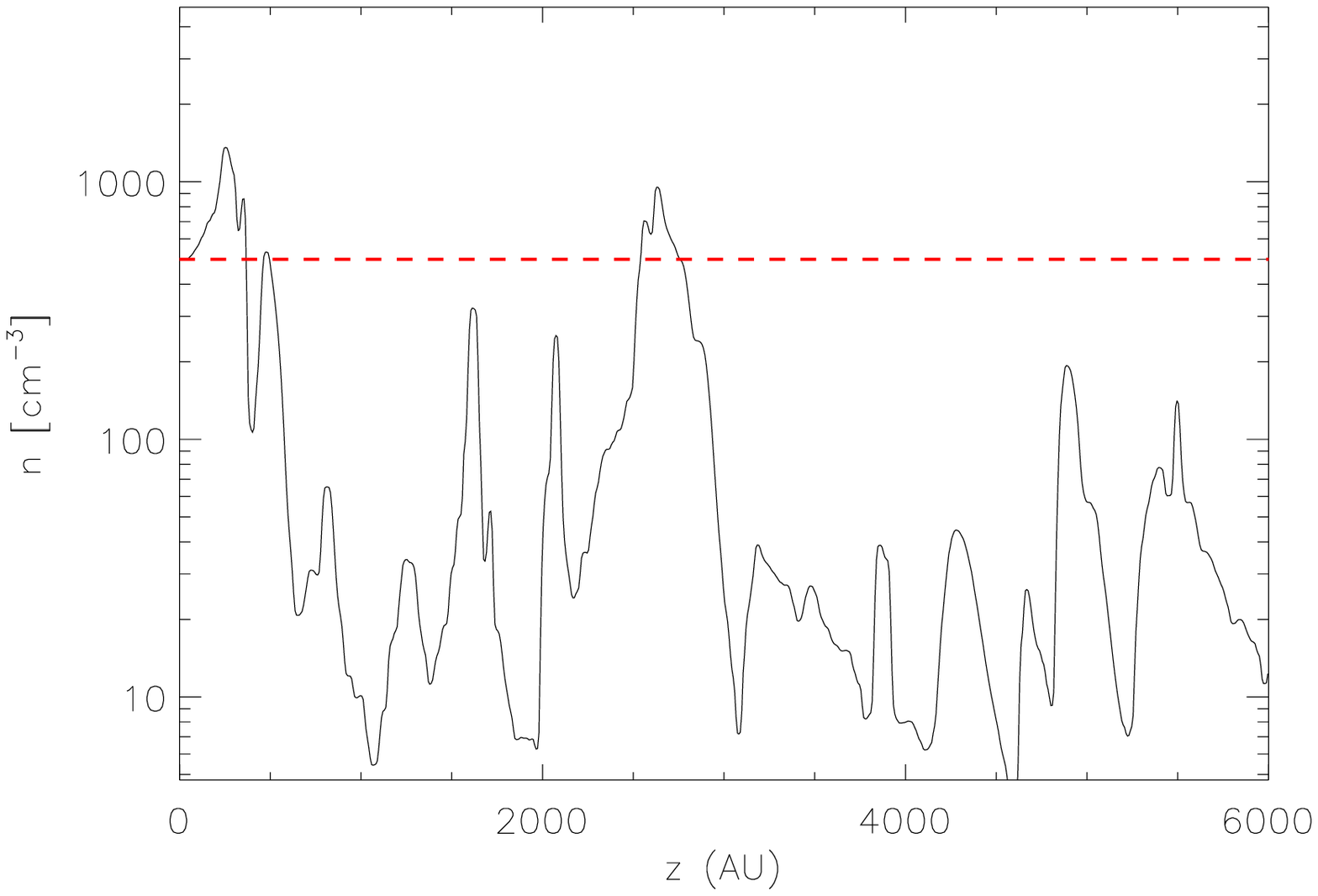,width=\columnwidth}}}
\centerline{\hbox{
\psfig{figure=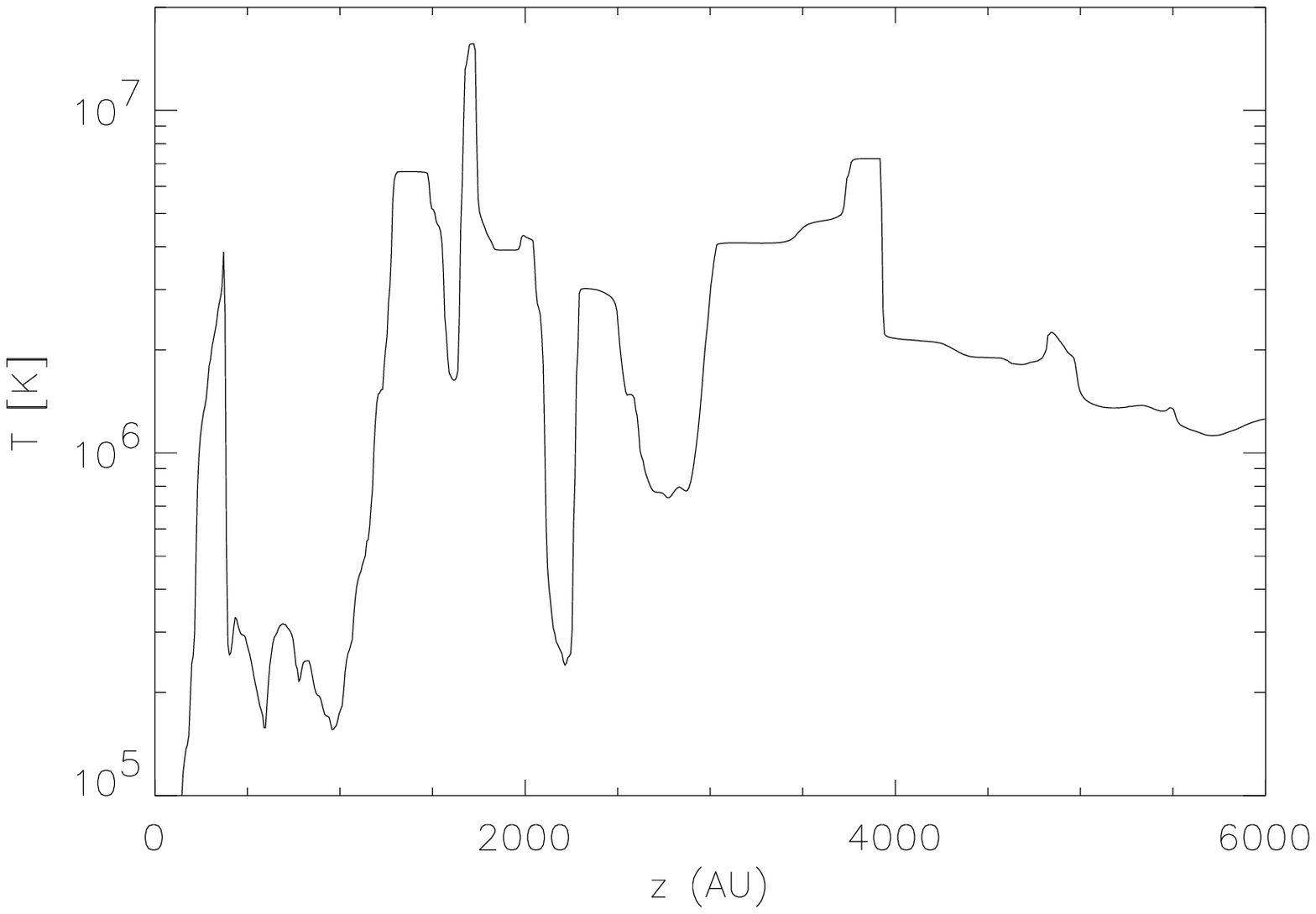,width=\columnwidth}}}
\caption{Cut along the jet axis of the spatial distributions of density
(upper panel) and of temperature (lower panel) in run LJ0.5: the
chain of knots within the jet due to previously ejected blobs lead to a
complex structure of the medium where the new blobs are ejected. The red
dashed line superimposed in the density profile indicates the initial jet
density value ($n_{\rm j} = 500$ cm$^{-3}$).}
\label{profilo-dens-T-05}
\end{figure}

In all our simulations, the jet propagates through the homogeneous
ambient medium, forming a cocoon, and shows a complex density structure
inside. The ejection rate determines the energy powering the jet and,
in fact, the maximum distance from the driving source reached at a given
epoch in runs LJ0.5, LJ2, and LJ8 strongly varies. As shown
in Fig. \ref{dens-2-8}, the low ejection rate case (LJ8) travels through
3000 AU in 100 yr, while the high ejection rate cases reach the 6000 AU
distance in less than 90 yr (LJ2) or even less than 50 yr (LJ0.5), being
the kinetic power released in these cases much higher than in LJ8.

Although the ambient medium is initially homogeneous, it becomes quickly
inhomogeneous as the first and consecutive blobs ejected by the stellar
source interact with it. Fig. \ref{profilo-dens-T-05} shows a
cut along the jet axis of the density (upper panel) and of the temperature
(lower panel) distributions. The whole computational domain (6000 AU) is
filled with several structures, with the density varying by more than two
orders of magnitude. Also the thermal conditions vary along the jet axis,
so that the sound speed and the Mach number are not uniform. As
a result, strong variations of the shock velocities ($v_{sh} \propto
\rho^{-1/2}$) and of the post-shock properties are determined due to the
varying pre-shock conditions. Note that, at variance with the model of a
continuous jet, in our pulsed jet model the density contrast between the
ambient medium and the blob can vary during the jet/ambient evolution. As
discussed above, in a few years, the medium is so complex that, even if
the initial density of the ejected blobs does not vary ($n_{\rm j} = 500$
cm$^{-3}$, see the red dashed line superimposed in the upper panel of
Fig. \ref{profilo-dens-T-05}), the ambient density strongly varies during
the evolution thus leading to several different scenarios in which the
ejected blob could be denser or less dense than the surrounding medium
(see upper panel in Fig.  \ref{profilo-dens-T-05}).

It is worth emphasizing that the random velocity of the blobs
ejected at different epochs together with the strongly inhomogeneous
medium in which each blob (except the first one) propagates, lead
to complex mutual interactions of the blobs. The frequency of mutual
interactions within the jet depends on the ejection rate, being the
largest in run LJ0.5 and the lowest in run LJ8. These interactions
result in a rich variety of plasma structures that cannot be described by
current models of jet and that determine an irregular pattern of knots.
In the following, we illustrate the common features that may contribute
in determining the morphology of the pulsed jet by analyzing the density
and temperature spatial distributions and evolution.

\begin{figure*}[!t]
\centerline{\psfig{figure=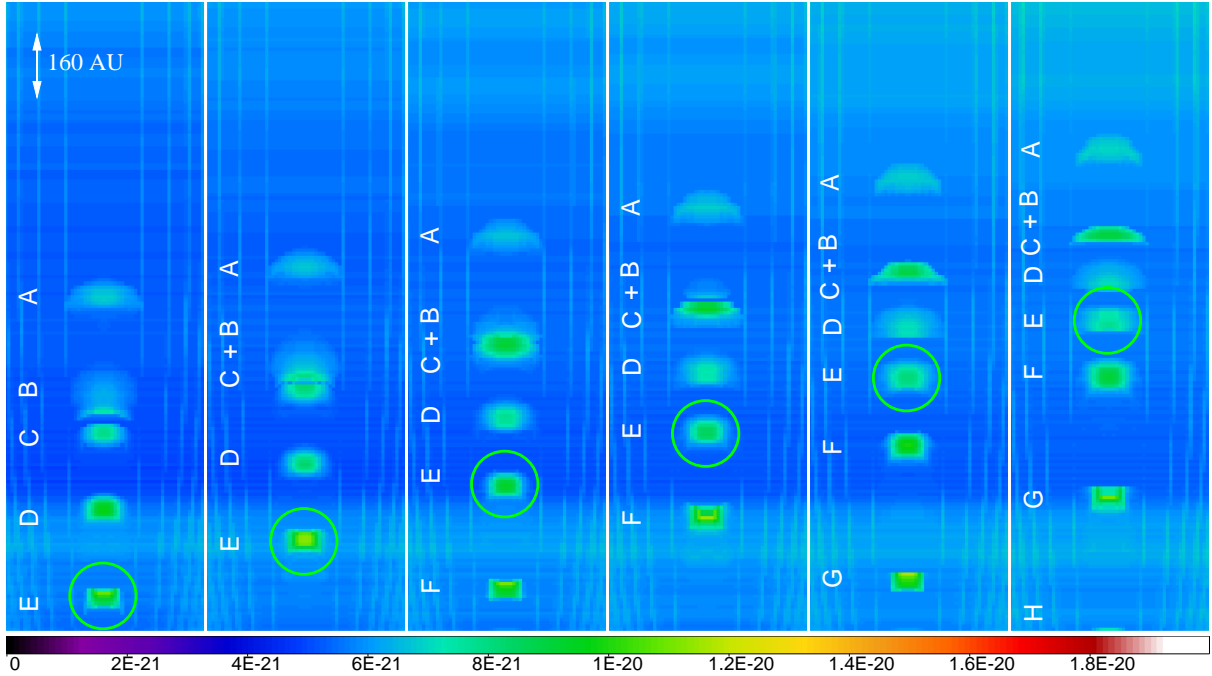,width=16cm}}
\caption{Density maps of plasma with temperature ranging between $(5000
- 100000)$ K (a proxy of optical emission) for the model HJ2. The green
circles (of radius $\approx 80$ AU) superimposed to each frame mark the
position of a specific knot as it evolves in time: the proper motion is
detectable in a few years, the time-lapse between two consecutive frames
is $1$ yr.}
\label{dens-T-ott-3D-intert-2}
\end{figure*}

The most evident feature predicted by our model is an irregular
chain of knots aligned along the jet axis. Depending on the model
parameters, the knots may emit mainly either in the optical band or in
the X-ray band. In the former case, the chain of (optical) knots can be
highlighted by deriving density maps of plasma with temperature ranging
between $(5000 - 100000)$ K, which is a proxy of optical emission. To
this end, we first reconstruct the 3D spatial distributions of mass
density and temperature, according to the symmetry of the problem; then,
assuming the jet traveling perpendicularly to the line-of-sight (LoS),
we integrate the density along the LoS and derive the 2D density maps. As
an example, Fig. \ref{dens-T-ott-3D-intert-2} shows an enlargement of the
base of the jet for run HJ2, i.e. in the heavy jet scenario (see
Tab. \ref{tab_mod}), where a chain of (optical emitting) knots within
the jet is well visible. These knots show a detectable proper
motion in a few years, in good agreement with typical observations of
the moving optical knots within protostellar jets. In particular, by
measuring the spatial displacement of the center of the circular region
identifying the E knot in Fig. \ref{dens-T-ott-3D-intert-2} in the time
interval $(45 - 50)$ yr, we derive an average knot speed along the
jet axis of about $600$ km/s, in nice agreement with the values obtained
from the optical observations of the knots within protostellar jets
(\citealt{fld05}; \citealt{bff08}, \citealt{emb94}).
In this case, we find evidence of
knotty structures analogous to those observed in protostellar jets,
with a chain of several subsequent knots in a few hundreds AU from
the stellar source.

\begin{figure}[!t]
\centerline{\psfig{figure=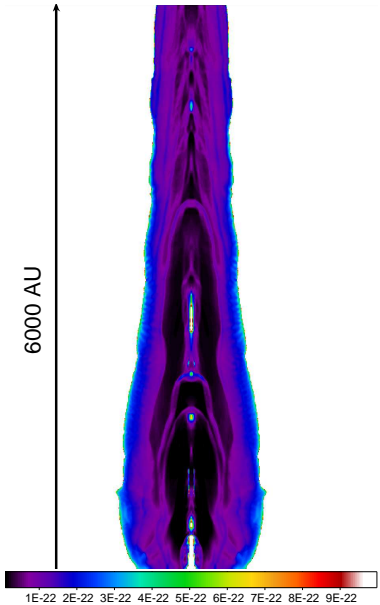,height=8cm}}
\caption{2D spatial distribution of the density of the jet/ambient
interaction in model LJ0.5: a chain of several knots arises due to blobs
ejected at different epochs.}
\label{dens-05}
\end{figure}

Similar patterns of knots but emitting mainly in the X-ray band (see
Bonito et al., 2009, in preparation) are found in the light jet scenario when the ejection rate is high ($\Delta t = 0.5, 2$ yr), as shown in the 2D spatial distribution of mass density in
Fig. \ref{dens-05}. In particular several dense structures are well
visible within the jet, mainly on the jet axis. 

The spatial separation between two consecutive knots is neither
uniform in space nor constant in time in all the different cases
analyzed because of the variation of the blob speed and of the medium
surrounding the blob. This leads to a complex pattern consistent with
observations of several protostellar jets, among them HH 154 (see
\citealt{bff08}). Sinusoidal ejection velocity models usually show a
chain of knots with an almost regular pattern (\citealt{rdk07}) within
the jet which does not reproduce the observed structure of complex jets,
(e.g. HH 1, \citealt{rhy00}; HH 111 \citealt{hmr01}).

\begin{figure}[t!]
\centerline{\hbox{
\psfig{figure=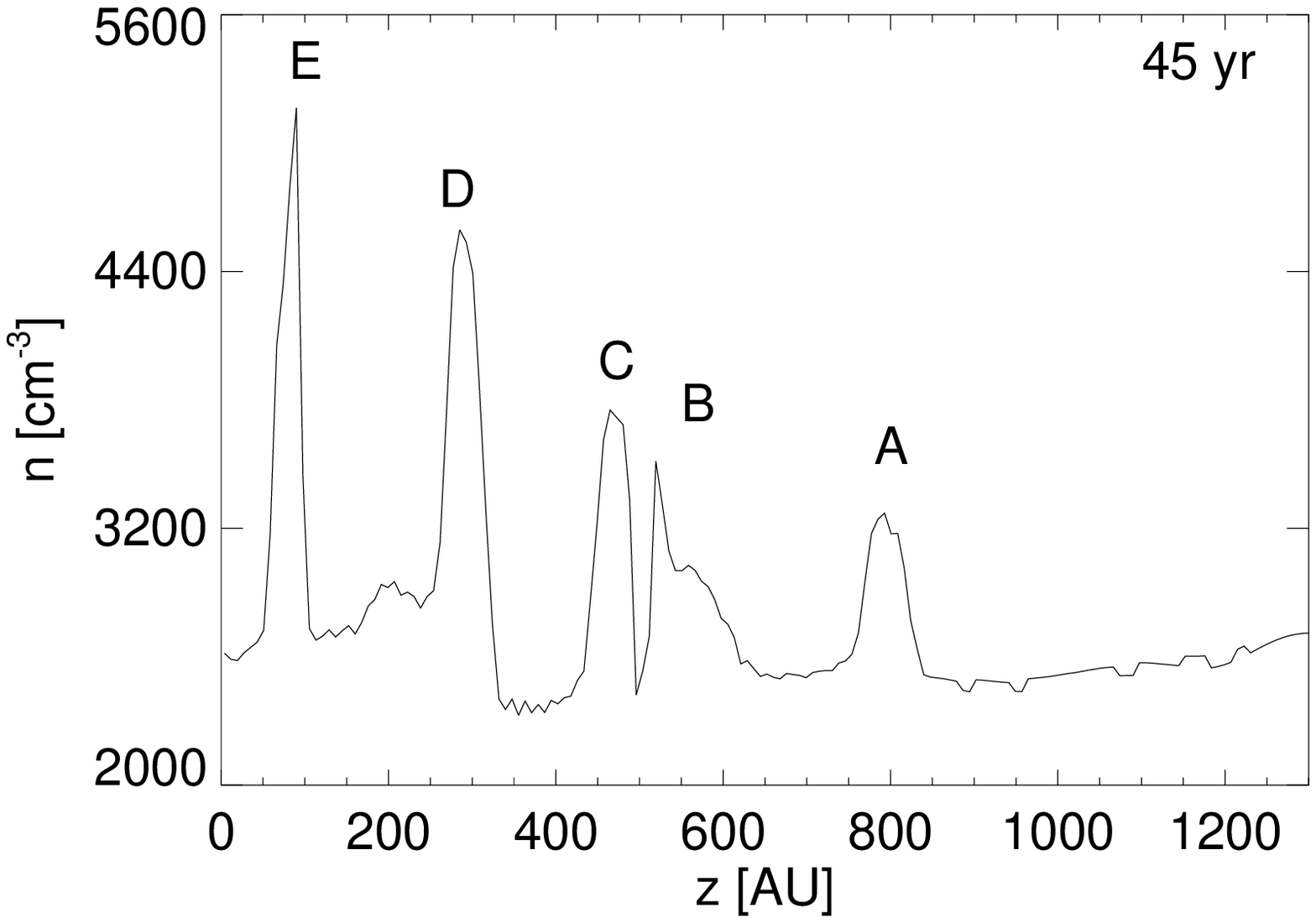,width=\columnwidth}}}
\centerline{\hbox{
\psfig{figure=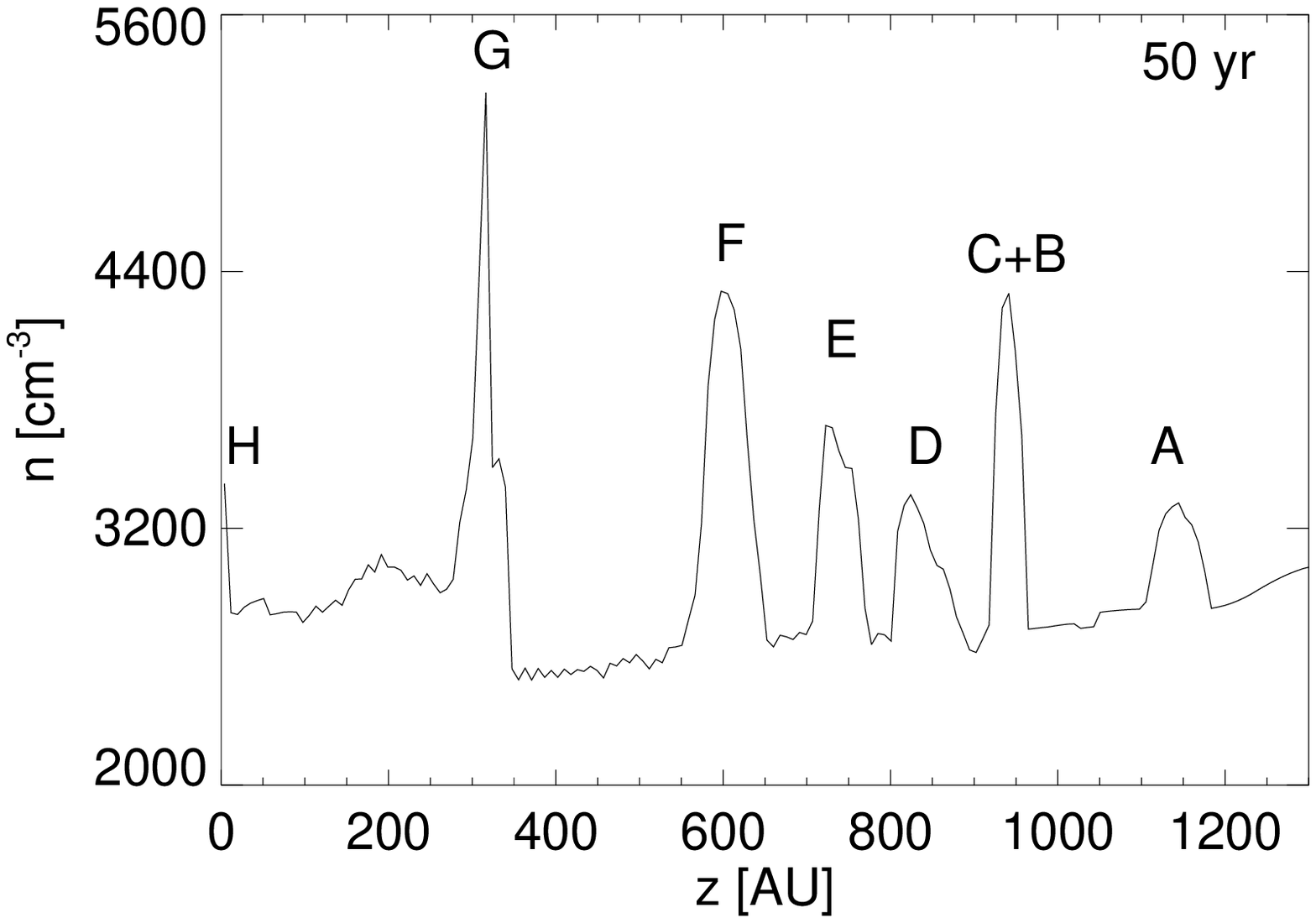,width=\columnwidth}}}
\caption{Cuts along the jet axis of the first (upper panel) and the last
(lower panel) frame of Fig. \ref{dens-T-ott-3D-intert-2}, corresponding to
45 yr and 50 yr since the beginning of the jet/ambient interaction. In a
few years, the proper motion of the knots within the jet, the interaction
between knots,
and the generation of new knots are well visible.}
\label{proj}
\end{figure}

The knots resulting from the interactions among ejected blobs have, in
general, different proper motions and may ultimately interact with each
other. In some cases, high speed knots can interact with the head
of the jet, possibly overtaking the cocoon, and thus traveling into
the unperturbed ambient medium. A random variability of the injection
velocity of the blobs and the consequent catching-up processes between
knots have been discussed previously by \citet{rag92}. This effect
is evident in Fig. \ref{dens-T-ott-3D-intert-2}, where the initially
separated $C$ and $B$ knots merge in a single structure (named $C+B$
in the figure). Analogously the density cut along the jet axis indicate
that the two peaks $C$ and $B$ at $t\approx 45$ yr of evolutionary stage
(upper panel of Fig. \ref{proj}), correspond to a single peak $C+B$
after $5$ yr (lower panel of Fig. \ref{proj}). These features have been
observed in several HH jets (\citealt{bff08}; \citealt{gef09}).

Analogous clump-clump interactions and high speed knots overtaking
the head of the jet (including also the case of off-axis structures) have
been found by \citet{yfc09}, although these authors have not considered
any proxy of optical emission as done here. Note, however, that the
knots in \citet{yfc09} are the result of imposed inhomogeneities of the
jet flow (describing the effect of possible instabilities) and their
characteristic size is lower than the jet diameter (subradius), whereas
in our model the knots are the result of the pulsed ejection of the jet
(possibly related to accretion phenomena) and their size is comparable
to the jet diameter. Note, also, that \citet{yfc09} considered ranges of
temperatures and velocities smaller than those discussed here and their
model does not take into account the thermal conduction effects. As
discussed below in Sect.~\ref{Thermal conduction}, our simulations show
that the thermal conduction plays a crucial role in damping out the
hydrodynamic instabilities that would develop within the jet, leading
to a faster and more collimated jet.

\begin{figure*}[!t]
\centerline{\psfig{figure=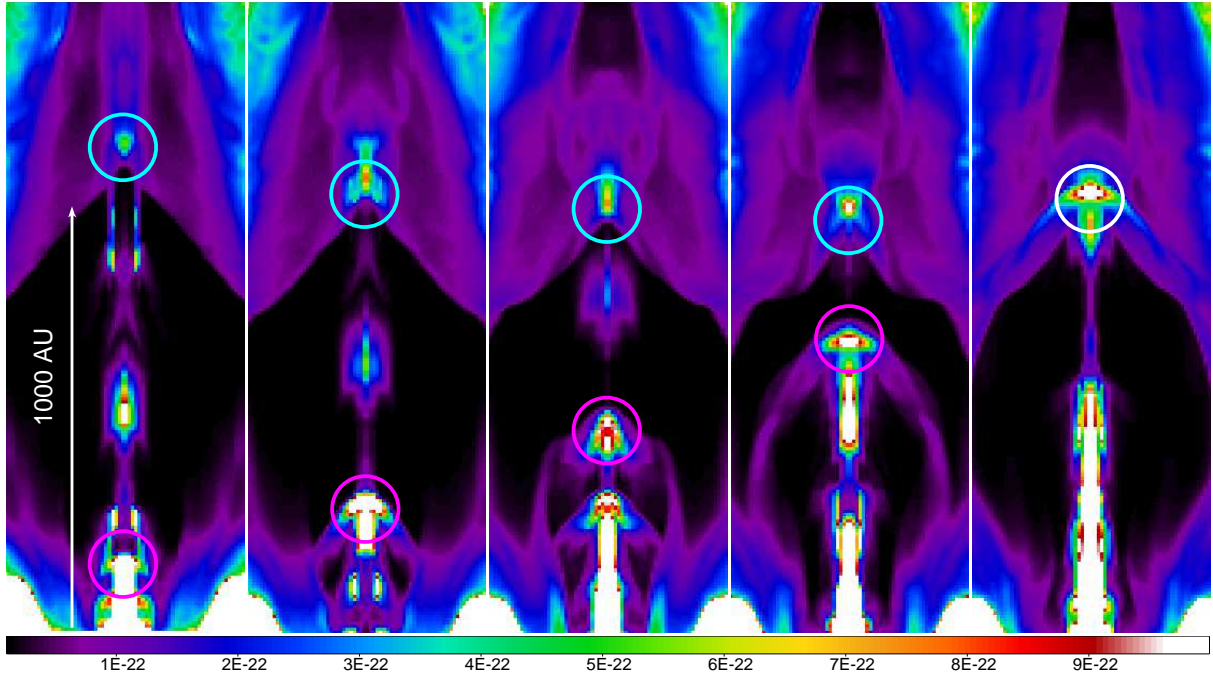,width=16cm}}
\caption{Evolution in a few years of the 2D spatial distributions of the
density of the jet/ambient interaction in model LJ0.5: reverse shocks
(highlighted by the cyan circle in each panel) form and travel in the
direction opposite to the ejected blobs (highlighted by the magenta circle
in each panel) and interact with them (white circle superimposed on the
last panel).}
\label{dens-05-2}
\end{figure*}

Another important feature produced by our model is associated with
the formation of reverse shocks moving in the opposite direction with
respect to the ejected blobs and interacting with these (see Fig. \ref{dens-05-2}). This interaction
produces, under certain circumstances, the formation of shocks with
no detectable proper motion over time scales of a few years, as observed,
for instance, in the X-ray source of HH 154 (\citealt{fbm06}).
The collision between knots and reverse shocks is a feature common
to all the ejection rates discussed here and occurs several times during
the time covered in our simulations. As an
example, Fig.  \ref{dens-05-2} shows one of these occurrence for run
LJ0.5. In the first panel, the reverse shock is located $\approx
1200$ AU above the source driving the jet and the knot traveling towards
it is initially at about $200$ AU from the source. As the jet evolves,
the knot reaches a distance of $\approx 800$ AU from the source,
with the reverse shock just $\approx 200$ AU above it. After about two
years the two structures interact at a distance of $\approx 1000$ AU from
the source and the shock location does not change appreciably over
a timescale of few years.

\begin{figure*}[!t]
\centerline{\psfig{figure=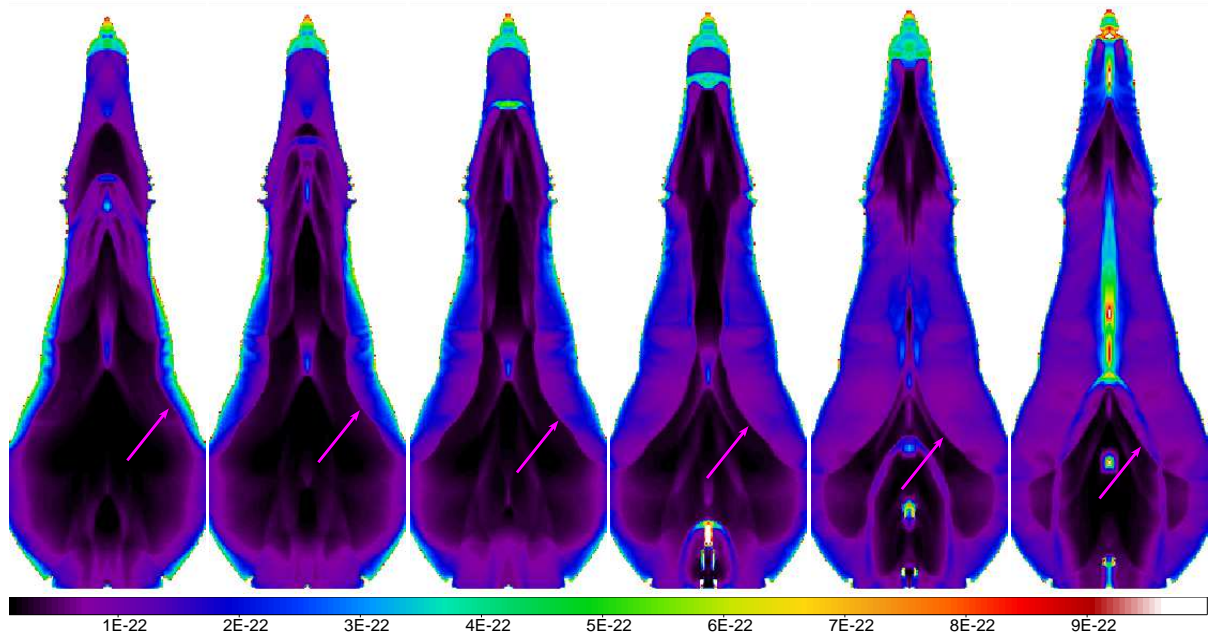,width=16cm}}
\caption{Model LJ8: evolution of the 2D spatial distributions of the
density of the jet/ambient interaction (the first frame is after $70$
yr and the last after $75$ yr since the beginning of the jet/ambient
interaction).  Oblique shocks form (marked by the arrows in each panel)
as the ejected blob interacts with the lateral cocoon.}
\label{rho-8}
\end{figure*}

A more complex morphology arises when the reverse shock is reflected
obliquely by the cocoon, leading, in this case, to an oblique shock
pattern. Fig. \ref{rho-8} shows the evolution of an oblique shock
in a few years, moving from the external cocoon toward the jet axis,
as indicated by the arrows superimposed on each frame. Note that
the propagation of the oblique shock causes the formation of a dense
and well collimated core within the jet itself as evident in the last
frame of Fig. \ref{rho-8}.

\subsection{Effects of the thermal conduction}
\label{Thermal conduction}

\begin{figure}[!t]
\centerline{\psfig{figure=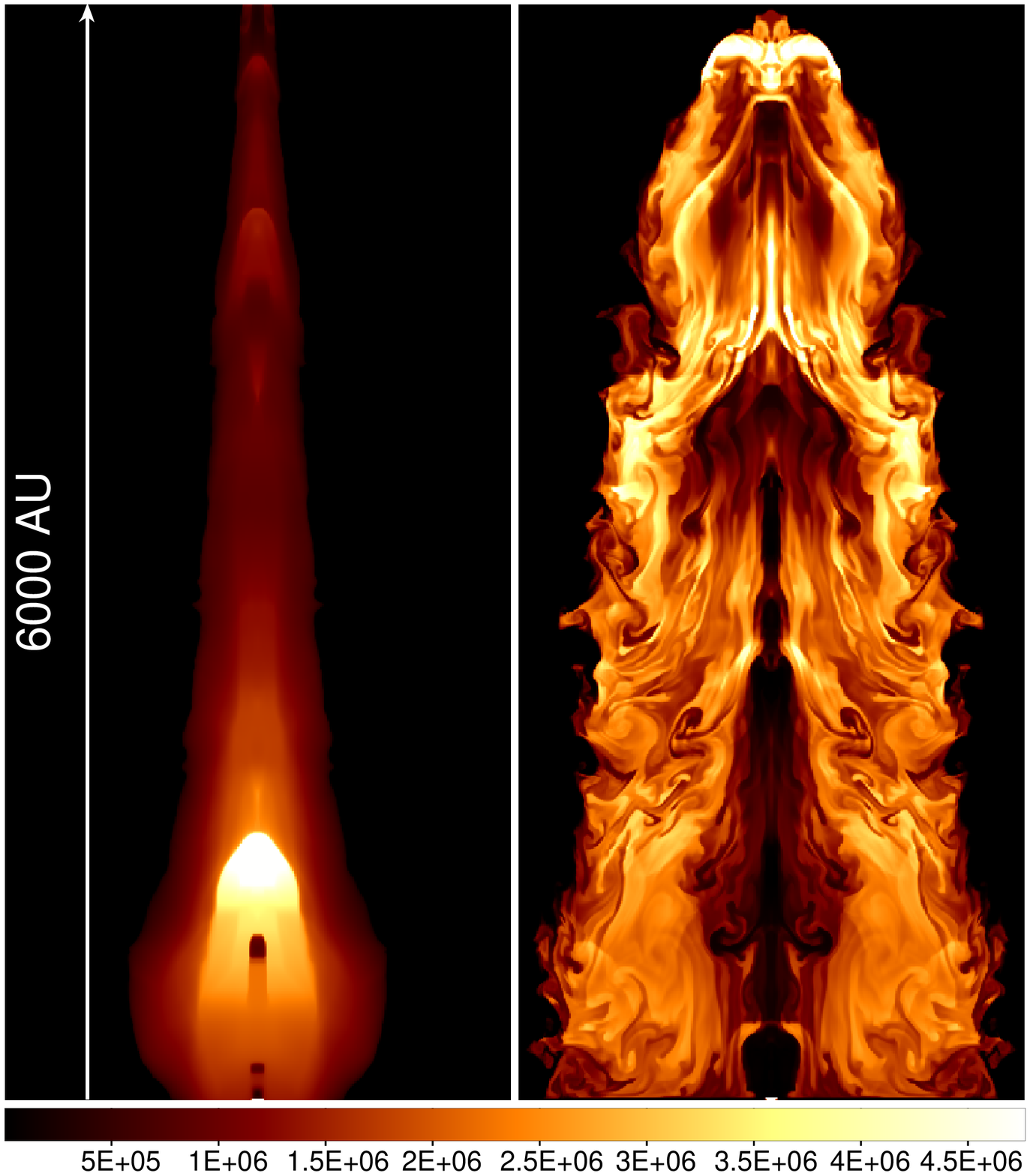,width=6cm}}
\caption{2D spatial distributions of the temperature of the jet/ambient
interaction in the $\Delta t = 2$ yr case: with radiative losses
and thermal conduction after about $90$ yr (left panel) and the pure
radiative case after about $350$ yr (right panel) of evolution. Ignoring
the effects due to the thermal conduction, the whole computational domain
is fully perturbed along the jet axis after more than $350$ yr.}
\label{T-intert-2-no-cond}
\end{figure}

Our model takes into account both the radiative losses from an optically
thin plasma and the thermal conduction, in Spitzer or in saturated
regimes. It is interesting, therefore, to investigate the effects of
the thermal conduction in the dynamics and energetic of the jet. These
effects are highlighted by comparing simulations either with or without
the thermal conduction. To this end, we performed an additional run
with setup identical to that used for run LJ2 but without the thermal
conduction (run LJ2RD). We find that, in run LJ2RD, the maximum distance
from the stellar source is significantly smaller than that
in run LJ2. As an example, Fig. \ref{T-intert-2-no-cond} shows the
temperature distributions in the two cases after the first blob has
reached the upper boundary of the computational domain, namely at
$t\approx 90$ yr in LJ2 and at $t\approx 350$ yr in LJ2RD. We found,
therefore, that the jet travels more slowly into the ambient medium if
the thermal conduction effects are inhibited. In fact, in LJ2RD, the
jet evolution is dominated by thermal and hydrodynamic instabilities
(see right panel in Fig. \ref{T-intert-2-no-cond}) that are known to
contribute to the jet breaking. Conversely, these instabilities are
damped out by the thermal conduction in run LJ2, and the jet propagates
to larger distances.

The comparison between the two temperature maps of runs LJ2 and LJ2RD
in Fig. \ref{T-intert-2-no-cond}
indicate also significant differences in the morphology due to the
presence of the thermal conduction. In fact, beside the different
evolutionary stages needed in the two cases in order to reach the top
of the computational domain, it
is also evident how the thermal conduction leads to a smooth and more
collimated spatial distribution of temperature. The radiative case
LJ2RD shows a significant radial spread of the temperature distribution,
accounting for the decelerated material traveling into the instabilities.

\section{Discussion}
\label{Discussion}

\subsection{Observable proper motion}
\label{motion}

As discussed in \citet{bop07}, the high initial jet velocity is not an
observable since we can measure just the speed of formed knots,
i.e. the velocity of the plasma structures formed by the interactions
among the blobs or of the blobs with the inhomogeneous surrounding
medium. In other words, we can observe the effect of jet/ambient
interaction, not the launching region itself. We are interested,
therefore, in deriving the average speed of the observable knots,
determining the internal density structure of the jet. To this
end, Fig. \ref{ave-vely} shows the density-weighted average of the
$z-$component of the plasma bulk velocity within a distance of $10$
pixels\footnote{Corresponding to a physical radius of $\approx 70$
AU.} from the jet axis for the three ejection rates analyzed
(runs LJ0.5, L2, L8). The velocity values are consistent with
typical knots speeds observed within optical jets (few hundreds km/s,
\citealt{em98}; \citealt{bfr03}; \citealt{bff08}). In particular we
find a very good agreement for the $\Delta t = 2$ yr case and the speed
value decreases with increasing $\Delta t$ parameter.

Following our model results, we can make predictions about the
initial ejection velocity leading to the observed speed. In order
to verify if our assumed values are reasonable, we can compare the
mechanical luminosity derived from our simulations with the observed
bolometric luminosity of the source of the HH 154 jet. In particular,
from the values of the initial velocity used here and shown
in Tab. \ref{tab_mod} and following the relation (18) in \citet{bop07},
the kinetic power derived from our model ranges between $L_{\rm mech}
\sim 10^{-7} - 10$ $L_{\odot}$, i.e. always below the observed bolometric
luminosity of the source of the HH $154$ jet, $L_{bol}\approx40$
$L_{\odot}$ (see \citealt{bop07}, Tab. 1).  Thus we conclude
that the initial blob velocity values assumed in our models lead to
reasonably small kinetic power.

Finally, following the relation (17) from \citet{bop07}, we derive
a jet mass loss rate ranging between $\dot{M}\sim10^{-11} - 10^{-9}$
$M_{\odot}$/yr. Once again, as discussed in details in \citet{bop07},
these results concerning both the kinetic power and the mass loss rate
are several orders of magnitude lower than those observed in CO outflows
and it seems reasonable that the simulated protostellar jets cannot
drive molecular outflow.

\subsection{Ejection rate of the blobs}

\begin{figure}[!t]
\centerline{\psfig{figure=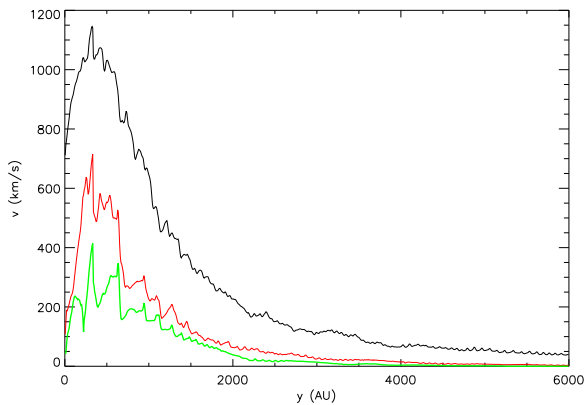,width=8cm}}
\caption{Average speed along the jet axis for the three runs LJ0.5, LJ2,
and LJ8: the $\Delta t = 0.5$ yr case is in black, the $\Delta t = 2$ yr
case is in red, and the $\Delta t = 8$ yr case is in green. The values
of the knots speed derived from the runs are consistent with optical
observations.}
\label{ave-vely}
\end{figure}

Our simulations show that observable knots are mainly produced by the mutual interactions among blobs
of plasma. We expect,
therefore, that the frequency of interactions between knots depends on
the ejection rate of the blobs (see Sect. \ref{The model}). In fact,
from our exploration of the parameter $\Delta t$, we find that the knot
interactions occur quite frequently in runs with high ejection rate (LJ05,
and LJ2), whereas it is almost absent in the low ejection rate case
(LJ8) over the timescale analyzed here. This result can be understood
from the following simple estimation.

\cite{bop07} showed that the observable velocity of a knot is a fraction of the initial jet velocity. This result is confirmed by our
simulations: as discussed in Sect. \ref{motion}, Fig. \ref{ave-vely}
shows that the maximum average speed of the knots along the jet axis is a fraction of the initial average blob velocity ($\approx 2000$ km s$^{-1}$).
The time at which we expect to observe a collision between two consecutive
knots is

\begin{equation}
t_{\rm coll} = \frac{v_{\rm knot}}{\Delta v_{\rm knot}} (\Delta t)_{\rm
knot}
\label{eq:tcoll}
\end{equation}

\noindent
where $\Delta v_{\rm knot}$ and $(\Delta t)_{\rm knot}$ are the average
relative speed and the time interval between two consecutive knots,
respectively. 
Both $v_{\rm knot}$ and $\Delta v_{\rm knot}$ are calculated assuming the same deceleration factor (since each knot velocity is a fraction of the initial blob speed); the latter is evaluated by considering only the blobs ejected with higher
speed with respect to the previous ones. 
Since knots are produced mainly by the mutual interaction between blobs and between blobs and the ambient, we expect that $(\Delta t)_{\rm
knot} \gsim \Delta t$. Considering $\Delta
t$ instead of $(\Delta t)_{\rm knot}$ in Eq.~\ref{eq:tcoll}, we have,
therefore, a lower limit on the collision time between knots. We derive
that the interactions rises on average over about $t_{\rm coll} = 10, 30$,
and more than 100 yr in runs LJ0.5, LJ2, and LJ8, respectively. This
estimate explains the presence of knot interactions in the high ejection
rate cases (LJ0.5 and LJ2) and the absence in the low ejection
rate case (LJ8) over the timescale considered. In the light
of the above discussion, we suggest, therefore, that the ejection rate
might be constrained if the frequency of mutual interactions among
ejected blobs can be determined from observations.
The constrain of the ejection rate could be also interesting in the context of the accretion scenario. In fact it is believed that the ejection/accretion phenomena are intimately related, but a full description of this relation is still lacking.

\section{Conclusions}
\label{Conclusions}

We investigated the physical mechanism determining the
irregular knotty structure of protostellar jets commonly observed
in the optical band, and recently also detected in X-rays (e.g. HH
154, \citealt{fbm06}). To this end, we developed a numerical model
describing a pulsed jet, consisting of a chain of blobs, each ejected along the jet axis with random initial velocity,
and traveling in an initial homogeneous ambient medium. Our simulations
represent the first attempt to model a pulsed jet that simultaneously
includes the thermal conduction and the radiative losses processes. In
particular, we find that the thermal conduction plays a crucial role
in damping out the hydrodynamic instabilities that would develop within
the cocoon. Since these instabilities contribute to break the jet,
the thermal conduction makes the jet to propagate to larger
distances.

The jet is modeled both as an initial light jet (a scenario which
seems likely to describe the observations of HH 154, both in X-rays and in
the optical band; \citealt{bff08}) and as an initial heavy jet. In all the
cases, we find that the interactions among subsequent ejected blobs are
a common feature to our simulations. As a result of these interactions,
each new ejected blob travels through a strongly inhomogeneous medium
characterized by density and thermal conditions largely varying along
the jet axis; as a result, the density contrast between the blob and
the surrounding medium varies during the jet/ambient evolution and the
different pre-shock conditions leads to strong variations of the shock
velocities and of the post-shock properties.

The mutual interactions among the blobs lead to a variety of
density structures within the jet not described by current models of jets (however see \citealt{yfc09}).
These structures determine a rather complex morphology of the jet. The
most relevant are: single high speed knots, showing a measurable proper
motion in nice agreement with optical and X-rays observations; irregular
chains of knots aligned along the jet axis, and possibly interacting
with each other, consistent with observations of several protostellar
jets; reverse shocks interacting with outgoing knots;
oblique shock patterns produced by the reflection of shocks at the
cocoon surrounding the jet. We also found that the range of velocity of
the knots formed from mutual interactions among blobs is consistent
with the typical range of values observed (e.g. \citealt{rb01}).

We explored the effects of the ejection rate of the blobs on the
evolution of the density and temperature spatial distributions of the
jet/ambient interaction. The frequency of interactions between
formed knots depends on the ejection rate. We suggest, therefore, that
the comparison between our model results and the observations may give
some hints on the ejection rate of protostellar jets.
Since the opposite and complementary processes of accretion of mass onto a star and ejection of mass from a star are commonly assumed to be related, from the constraints on the ejection rate we may also infer important physical information on the accretion rates. 

We expect some details of our model to depend on the assumption of
the exponential distribution for the ejection velocity of each blob
(determined with a given random seed).  From a quantitatively point
of view, the single blob velocity can change with a different
random seed and also the energetic and dynamics of particular structures
is expected to vary if different velocity distributions are chosen.
Nevertheless, in any case we expect to obtain the main features
discussed here, namely the formation of shock fronts, the collision
among knots due to blobs ejected at different epochs with different
speed, the interaction among knots and reverse shocks.

The model of a randomly ejected pulsed jet is, therefore, a promising
scenario to explain the physical mechanism leading to the irregular knotty
structure observed within protostellar jets in a wide range of
wavelengths, from optical to X-rays. Further development of this work
will be reported in a forthcoming paper (Bonito et al., 2009, in preparation) where we will
investigate the morphology and variability of X-ray emitting structures
formed within the randomly ejected pulsed jet.

\begin{acknowledgements}

We would like to thank the referee, Dr. A. C. Raga, for his helpful comments and suggestions.
R.B., S.O., M.M, and G.P. acknowledge support by the Marie Curie
Fellowship Contract No. MTKD-CT-2005-029768.  The software used in this
work was in part developed by the DOE-supported ASC/Alliances Center
for Astrophysical Thermonuclear Flashes at the University of Chicago,
using modules for thermal conduction and optically thin radiation
constructed at the Osservatorio Astronomico di Palermo. The calculations
were performed at CINECA (Bologna, Italy) and on the cluster at the SCAN
(Sistema di Calcolo per l'Astrofisica Numerica) facility of the INAF --
Osservatorio Astronomico di Palermo.

\end{acknowledgements}

\bibliographystyle{aa}

\end{document}